\documentclass[twocolumn,superscriptaddress,amsmath,amssymb,pre]{revtex4}
\usepackage{graphicx}
\usepackage{dcolumn}
\usepackage{bm}
\usepackage{color}

\begin{document}

\title{Three-dimensional aspects of fluid flows in channels. I. Meniscus and Thin Film regimes}

\author{\firstname{R.} \surname{Ledesma-Aguilar}}%
\email{rodrigo@ecm.ub.es}
\affiliation{Departament d'Estructura i Constituents de la
  Mat\`eria. Universitat de Barcelona, 
Avinguda Diagonal 647, E-08028 Barcelona, Spain}
\author{\firstname{A.} \surname{Hern\'andez-Machado}} 
\affiliation{Departament d'Estructura i Constituents de la
  Mat\`eria. Universitat de Barcelona, 
Avinguda Diagonal 647, E-08028 Barcelona, Spain}
\author{\firstname{I.} \surname{Pagonabarraga}}
\affiliation{Departament de F\'isica Fonamental.  Universitat de
  Barcelona, Avinguda Diagonal 647, E-08028 Barcelona, Spain}
\date{\today}
\begin{abstract}
We study the forced displacement of a fluid-fluid interface in a three-dimensional channel formed 
by two parallel solid plates. Using a Lattice-Boltzmann method, we study situations in which a slip 
velocity arises from diffusion effects near the contact line.  The difference between the slip and channel 
velocities determines whether the interface advances as a meniscus or a thin film of fluid is left
adhered to the plates.  We find that this effect is controlled by the capillary and P\'eclet
numbers.   We estimate the crossover from a meniscus to a thin
film and find good agreement with numerical results.   
The penetration regime is examined in the steady state. We find that the occupation fraction of 
the advancing finger relative to the channel thickness is controlled by the capillary number and the viscosity 
contrast between the fluids.  For high viscosity contrast, Lattice-Boltzmann results agree with 
previous results. For zero viscosity contrast, we observe remarkably narrow fingers. The shape of the finger 
is found to be universal. 
\end{abstract}
\maketitle
\section{Introduction}
Advancing fronts in fluid systems involve the motion of a fluid-fluid interface, a surface that lives 
in a three-dimensional world, and which is often constrained by a solid boundary.  A typical example 
is that of an interface moving in a channel\cite{deGennes02,couder,pelce}. 

Examples of advancing fronts in channels are imbibition, a process in which a wetting fluid invades the 
channel due to an uncompensated capillary pressure, and the viscous fingering process\cite{couder,pelce,saffman}, 
where a low-viscosity(or high-density) fluid penetrates a high-viscosity(or low-density) one. 

The problem of viscous fingering in a channel has been widely studied in the framework of Hele-Shaw theory.  
A Hele-Shaw cell is the two-dimensional limiting case of a very thin channel, where the equations of motion are 
averaged over the channel thickness.  This reduces the interface to a line, the \emph{leading interface},  
that lives in the plane of the cell. As a consequence, the approximation discards any effects arising from the
full three-dimensional structure of the interface.

Nonetheless, penetration in the gap of a Hele-Shaw cell is a fundamental three-dimensional 
effect that has important repercussions in the viscous fingering problem.  
As theoretical studies have pointed out\cite{parkhomsy}, a thin film of viscous fluid left adhered 
to the cell plates as the front advances modifies the capillary pressure at the leading interface, thus altering 
the front morphology.  This has been confirmed in experiments of steady viscous fingers\cite{libchaber}, 
where the presence of a thin film led to fingers not predicted by the two-dimensional theory.  In the following paper,
we will address the role of the thin film in viscous fingers.

A thin wetting film is not the only consequence of a three-dimensional interfacial structure.   
In the context of liquid films spreading over dry substrates\cite{Troian01,Brenner01}, where 
a two-dimensional approximation is typically applied, three-dimensional effects are also important. For 
instance, the stability of a spreading front depends on the wetting properties of the 
fluid.  Experimentally, it has been observed\cite{Chuang02} that a crossover from unstable to 
stable fronts occurs when the dynamic contact angle exceeds $\pi/2$, a situation which renders 
the velocity field within the film three-dimensional.   

The problem of a moving interface in a three-dimensional channel must take into account 
a dynamic contact line, the intersection point between the 
fluid-fluid interface and the channel walls.  In classic fluid mechanics, a moving contact line
violates the usual no-stick boundary condition, leading to a divergent viscous dissipation\cite{deGennes01}.
Hence, contact line dynamics must consider a regularizing mechanism of the viscous dissipation singularity.  
A slip velocity in the vicinity of a driven contact line arises naturally in diffuse 
interface models of binary fluids\cite{Jacqmin01}, regularizing the singularity.  
These models consist of the usual Navier-Stokes equations coupled to a convection-diffusion equation 
of an order parameter\cite{bray}.  
Diffuse interface effects enter the force balance equations in the shape of order parameter 
gradients that play the role of a Young force.  
As a result, the contact line slips over the solid boundary\cite{Jasnow01,Jacqmin01}.  
Away from the contact line, order parameter gradients vanish 
and the stick boundary condition is recovered.  The size of the diffusion region, $l_D$, is then a measure of 
how strong slip is for a given system and is clearly an important parameter.  This size was estimated by Briant and 
Yeomans\cite{Yeomans01}, who characterize $l_D$ for the case of an interface subjected to shearing walls.  They focused on 
the dependence of $l_D$($L$ in their notation) on the model parameters, finding a scaling relation that was 
verified numerically.

Important implications arising from a relatively large or small slip velocity compared 
to the leading interface velocity in forced fronts can be foreseen.  Whenever both velocities 
are comparable, the interface should maintain a meniscus shape.  Conversely, as the slip 
velocity becomes small compared to the channel velocity the interface shape should 
develop as a finger, leaving a thin film of fluid adhered to the walls of the channel. 

In this paper we study the penetration process across the channel thickness.  
We study the motion of the full three-dimensional 
interface between two viscous fluids when it is subjected to a 
gravitational body force.  We treat the case of a strictly flat leading interface, focusing only in the 
details that pertain to the channel thickness. We work with symmetric 
fluids as well as with fluids of different densities or viscosities. 

We focus on two principal matters.  We first describe how the contact line and leading 
interface velocities are related, and propose the mechanisms that determine the velocity 
ratio.  We find that the velocity ratio is controlled by the force balance
at the interface and by diffusion effects localized at the contact line.

Secondly, we study the thin film that forms inevitably in the case of small slip.
In that case the front decouples from the contact line leading to the growth of a finger,
even when the interface is linearly stable to the Rayleigh-Taylor instability.
We find that the fraction of occupation of the thin film relative to the channel thickness 
is a function of the capillary number and the 
viscosity contrast between the fluids.  The high viscosity contrast case is validated by comparing our results to 	
the numerical work of Halpern and Gaver\cite{Halpern01} which is consistent with previous results
of Taylor\cite{Taylor01}.   For fluids with zero viscosity contrast, it turns out that the finger width 
has much lower values than for the high viscosity contrast case at fixed capillary number. 

The morphology of our fingers is very much alike to the Saffman-Taylor finger shape, a prediction of
the Hele-Shaw theory.  Nevertheless, it is important to stress that although the case of a flat 
leading interface is two-dimensional, the equations of motion are not equivalent to those of 
the Saffman-Taylor problem.  Therefore, penetration in the channel thickness cannot be attributed to the 
Saffman-Taylor instability.   Likewise, the selection rule of the steady state, \emph{i.e.}, 
the actual dependence of the thin film thickness with the front velocity, cannot 
be mapped to the theoretical predictions of the viscous fingering theory.   
 
We will address these matters by means of numerical simulations of the mesoscopic equations 
of the system. To do so, we take advantage of a powerful integration algorithm in fluid dynamics, 
the Lattice-Boltzmann scheme for binary fluids. 

The paper is organized as follows.  In Sec.~\ref{sec:Model} we present the equations that govern the system in the mesoscopic regime.  
In Sec.~\ref{sec:LB} we briefly present the Lattice-Boltzmann algorithm for binary
fluid flows.  Sec.~\ref{sec:Diff} is dedicated to simulation results of the forced interface,
 from which two steady state regimes are found; a non-penetrating regime, in which the 
interface advances as a meniscus, and a penetrating one, in which a single finger emerges and achieves
steady state. In Sec.~\ref{sec:Disc} we present a scaling argument of the equations of motion 
that leads to an estimate of the ratio between the slip and front velocities. Such argument explains the
crossover from one regime to the
other. In Sec.~\ref{subsec:asymm}
we extend our results to fluids of different densities or viscosities. Sec.~\ref{subsec:finger} is devoted to the steady state finger. 
Finally, in~Sec.~\ref{sec:Conclusions} we present the conclusions of this work.

\section{Governing Equations}
\label{sec:Model}
\begin{figure}[b!]
\centering
\includegraphics[width=0.45\textwidth]{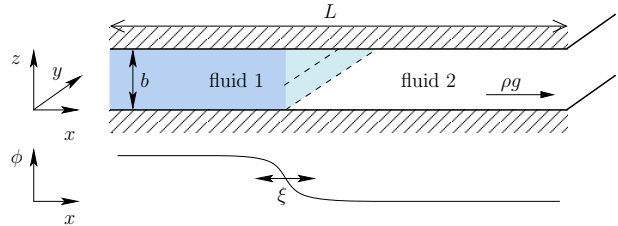}
\caption{Schematic representation of the system. The leading interface and 
contact line positions are indicated.  \label{fig:diagram}}
\end{figure}
We consider a channel formed by two solid plates parallel to the $xy$ plane, each 
of length $L$ and infinite width, located at positions $z=0$ and $z=b$. Initially, two fluids 
fill the channel and are separated by a flat interface perpendicular to the solid walls, as shown 
in Fig.~\ref{fig:diagram}.  The equilibrium contact angle is hence $\theta_E=\pi/2$. Contact lines 
are located at $z=0$ and $z=b$, while the leading interface is located at $z=b/2$. 

To circumvent the complications of the sharp interface formulation, we introduce a mesoscopic 
variable, $\phi(\vec r)$; and order parameter which is constant in the bulk of each fluid and 
varies smoothly across a diffuse interfacial region.  Within this approximation, the equilibrium state 
of the system is described by a Helmholtz free energy\cite{bray} 
$$\mathcal F\{\rho,\phi\}=\int\mathrm d \vec r  \left(V(\phi,\rho) + \frac{\kappa}{2}(\vec\nabla\phi)^2\right).$$ 
The first term in the integrand is a volume contribution, given by $V(\phi,\rho)=A\phi^2/2+B\phi^4/4 + \rho/3\ln\rho$.  
The $\rho$ dependent term corresponds to an ideal gas contribution, while  the $\phi$ dependent terms allow for the coexistence 
of two phases. The presence of an interface is accounted for by the last term in the integrand, which penalizes  
spatial variations of the order parameter by a factor $\kappa$. Minimization of $\mathcal F$ leads to the 
chemical potential, 
$$\mu = \partial_\phi V - \kappa \nabla^2\phi,$$
and total pressure tensor\cite{pagonabarraga}, 
$$\begin{array}{ccl}
\bar{\bar P}_T&=&\left( \frac{\rho}{3}+\phi\partial_\phi V - V -\kappa\left(\phi \nabla ^2\phi +\frac{1}{2}|\vec \nabla \phi|^2 \right)\right)\bar{\bar \delta}\cr
&& +\kappa\vec\nabla\phi\vec\nabla\phi,
\end{array}$$
where $\bar{\bar \delta}$ is the diagonal matrix.  The pressure tensor has an ideal contribution given by  
$\bar{\bar P}=\frac{\rho}{3}\bar{\bar \delta}$, and an order parameter contribution.  
In equilibrium, the order parameter profile for the flat interface~(sketched in 
Fig.~\ref{fig:diagram}) is $\phi^*(x,z)=-\phi_{eq}\tanh(x/\xi)$, where $\phi_{eq}=(-A/B)^{1/2}$ is the bulk 
equilibrium value of the order parameter and $\xi=(-\kappa/2A)^{1/2}$ is the length scale of the interfacial 
region; this profile leads to the difference between equilibrium values $\Delta \phi = 2\phi_{eq}$ and 
the energy per unit area of the interface, $\sigma=(-8\kappa A^3/9B^2)^{1/2}$.  Since the interface is diffuse, 
a choice for the \textit{nominal} interface position has to be made.  
We choose the level surface $\phi=0$.  
 
The divergence of the pressure tensor yields the force per unit volume that acts on the fluid:  $-\vec \nabla P - 
\phi \vec \nabla \mu$.  The first term is the pressure gradient, while the second arises from 
order parameter inhomogeneities.  Consequently, the Navier-Stokes equations are\cite{bray},
\begin{equation}
\rho \left(\mathrm \partial_t\vec v + \vec v \cdot \vec \nabla \vec v\right)  = - \vec \nabla P -\phi\vec\nabla\mu + \eta \nabla^2 \vec v + \rho \vec g, 
\label{eq:ns1}
\end{equation} 
where $\vec v$ is the fluid velocity,  $\eta$ is the fluid viscosity and $\vec g$ is the acceleration of gravity.  

The dynamics of the order parameter are described by a convection-diffusion equation, 
\begin{equation}
\partial_t\phi + \vec v \cdot \vec \nabla \phi =  M \nabla^2\mu,
\label{eq:cd1}
\end{equation}    
where $M$ is a mobility. For small deviations from the equilibrium configuration, 
an expansion of the chemical potential in powers of $\phi -\phi^*$ yields a  first order 
diffusion coefficient $D=M(A+B\phi_{eq}^2)$, so  the relative importance of the 
advective and diffusive terms can be estimated through a P\'eclet number, 
$Pe = |\vec v \cdot \vec \nabla \phi|/|D \nabla^2\phi|$. 

The system can be represented as a sheet of fluid in the $xz$ plane with periodic boundary 
conditions applied in the $y$ direction. This is equivalent to a channel of infinite width 
in the $y$ direction with a flat leading interface.  Stick boundary conditions are imposed 
at the walls, $\vec v(x,z=0)= \vec v(x,z=b)=\vec 0$, while no flow boundary conditions are imposed 
for the order parameter, $\phi \vec v(x,z=0)= \phi \vec v(x,z=b)=\vec 0$.  At both ends of the channel 
the flow is homogeneous. This is ensured by setting $\partial_x \rho \vec v (x=0,z) = 
\partial_x \rho \vec v (x=L,z) = \vec 0$ and $\partial_x \phi \vec v (x=0,z) =
\partial_x \phi \vec v (x=L,z) = \vec 0.$

Contact line dynamics arise from the diffuse nature of the interface, which allows for slip in  
the interfacial region by a diffusive mechanism. The size over which slip takes place, $l_D$,  
is a function of the fluid properties and has been estimated by Briant and Yeomans\cite{Yeomans01}, who  
have given a scaling relation, $l_D\sim(\eta \xi^2M/\Delta \phi^2)^{1/4}$. 

\section{Lattice Boltzmann Method} 
\label{sec:LB}
We solve numerically Eqs.~(\ref{eq:ns1}) and (\ref{eq:cd1}) by means of the Lattice-Boltzmann algorithm 
presented in Ref.~\cite{pagonabarraga}.  The dynamics are introduced by discretized Boltzmann equations 
of two distribution functions,
\begin{equation}
f_i(\vec r + \vec c_i,t + 1) - f_i(\vec r,t) = -\frac{1}{\tau_f}(f_i - f_i^{eq}) + F_i^f,
\label{eq:evf}
\end{equation}
and 
\begin{equation}
g_i(\vec r + \vec c_i,t + 1) - g_i(\vec r,t) = -\frac{1}{\tau_g}(g_i - g_i^{eq}).
\label{eq:evg}
\end{equation}
In these equations, $f_i$ and $g_i$ are distribution functions, 
where the index $i$ counts over the model velocity set.  Space is discretized as a cubic lattice where nodes 
are joined by velocity vectors, $\vec c_i$.  Space and time units in Eqs.~(\ref{eq:evf}) and~(\ref{eq:evg}) are set to unity.  
Likewise, the density of the fluids is set to one.   We use the D3Q15 velocity set, which consists of fifteen
velocity vectors:  six of magnitude 1 that correspond to nearest neighbors, eight of magnitude $\sqrt{3}$ that
correspond to third-nearest neighbors and one of zero magnitude
that accounts for rest particles. In the D3Q15 model the speed of sound is $c_s=1/\sqrt{3}$.  
In Eqs.~(\ref{eq:evf}) and~(\ref{eq:evg}), distribution functions are first relaxed to equilibrium values, 
represented by $f_i^{eq}$ and $g_i^{eq}$, with relaxation timescales $\tau_f$ and $\tau_g$.  The term $F_i^f$ is related 
to the external forcing.  Following the collision stage, distribution functions are propagated to neighboring sites.
 
Hydrodynamic variables are defined through moments of the $f_i$ and $g_i$. The local density 
and order parameter are given by $\sum_if_i=\rho$ and $\sum_i g_i=\phi.$   The fluid  momentum and order parameter current, 
are defined as $\sum_if_i \vec c_{i} =\rho \vec v $ and $\sum_ig_i \vec c_{i} =\phi \vec v $.  Local conservation
of mass and momentum is enforced through the conditions $\sum_if^{eq}_i  =\rho $, $\sum_ig^{eq}_i =\phi$,
$\sum_if^{eq}_i \vec c_{i} =\rho \vec v $ and $\sum_ig^{eq}_i \vec c_{i} =\phi \vec v $. 
In equilibrium, the pressure tensor and chemical potential are defined as $\sum_if^{eq}_i\ \vec c_{i}\vec
c_{i}  =\rho \vec v \vec v + \bar{\bar P}_T$ and $\sum_ig^{eq}_i\ \vec c_i\vec c_i  =\hat M \mu\bar{\bar
\delta} + \phi \vec v \vec v$.

The equilibrium distribution functions and the forcing term are written as expansions in powers of $\vec v$\cite{ladd}, \textit{i.e.},  
$$f_i^{eq} = \rho \omega_\nu\left(A_\nu^f + 3\vec v\cdot \vec c_{i} + \frac{9}{2}\vec v\vec v : \vec c_{i}\vec c_{i} -\frac{3}{2}v^2 + {\bar{\bar G}}^f:\vec c_{i} \vec c_{i}\right),$$
$$g_i^{eq} = \rho \omega_\nu\left(A_\nu^g + 3\vec v\cdot \vec c_{i} + \frac{9}{2}\vec v\vec v : \vec c_{i}\vec c_{i} -\frac{3}{2}v^2 + {\bar{\bar G}}^g:\vec c_{i} \vec c_{i}\right)$$
and 
$$F^f_i = 4\omega_\nu\left(1-\frac{1}{2\tau_f}\right)\left[\vec f \cdot \vec c_i(1+\vec v \cdot \vec c_i)-\vec v \cdot \vec f\right].$$
Here, $\nu$ stands for the three possible magnitudes of the $\vec c_i$ set.  Coefficient values are 
$\omega_0 = 2/9,$ $\omega_1=1/9$ and $\omega_{\sqrt{3}}=1/72;$ $A^f_0=9/2-7/2\mathrm{Tr}\bar{\bar P} ,$
$A^f_1 = A^f_{\sqrt{3}} = 1/\rho \mathrm{Tr}\bar{\bar P}$ and $\bar{\bar G}^f = 9/(2\rho)\bar{\bar P}-
3\bar{\bar \delta}\mathrm{Tr}\bar{\bar P};$  $A_0^g= 9/2-21/2\hat M\mu,$  $A_1^g = A^g_{\sqrt{3}} = 3\hat M\mu/\rho$ and 
$\bar{\bar G}^g = 9/(2\rho)\hat M\mu(\bar{\bar1}-\bar{\bar\delta}),$ where $\bar{\bar 1}$ is the unit matrix.  

Eqs.~(\ref{eq:ns1}) and~(\ref{eq:cd1}) can be recovered as a Chapman-Enskog expansion 
of Eqs.~(\ref{eq:evf}) and~(\ref{eq:evg})\cite{ladd}.  The Lattice-Boltzmann scheme maps 
to the hydrodynamic model through the relaxation timescales, \textit{i.e.}, $\eta=(2\tau_f - 1)/6$ 
and  $M=(\tau_g - 1/2)\hat M,$ and through the body force $\vec f = \rho \vec g$.

Solid boundaries in the Lattice-Boltzmann method are implemented by means of the well known bounce-back 
rules\cite{ladd,Pagonabarraga02}.  In the lattice nodes that touch the solid, the propagation scheme is
modified so the distribution functions are reflected to the fluid rather than absorbed by the solid.  As a
consequence, a stick condition for the velocity is recovered approximately halfway from the fluid node to the
solid node.
 
\section{Results} 
We study the process of penetration across the channel thickness in the presence of a dynamic contact line. 
As we have explained above, fingering is expected whenever the slip velocity is small compared to the leading 
interface velocity. In our model, slip is controlled by diffusion in the vicinity 
of the contact line.  To measure the importance of diffusivity we use a typical definition of the P\'eclet number, 
$Pe=Ub/D$, where $U$ is the velocity of the leading interface.  The other relevant control parameter 
is the capillary number, which follows from the ratio between viscous and capillary forces, $Ca=\eta U/\sigma$.
We focus on flows governed by viscous and capillary forces.  To enforce this situation we neglect the convective 
term in Eq.~(\ref{eq:ns1}). To assure that we work on the low Mach number regime, the fluid velocity is restricted to $U\leq 0.01$. 
For the case of small slip, we expect a thin film regime typical of 
experiments.  We characterize this regime in terms of the finger 
width, viscosity contrast and capillary number.  We compare our results with 
other studies from the literature.     

\subsection{Effect of diffusivity, surface tension and viscosity}
\label{sec:Diff}
\begin{figure}
\includegraphics[width=0.45\textwidth]{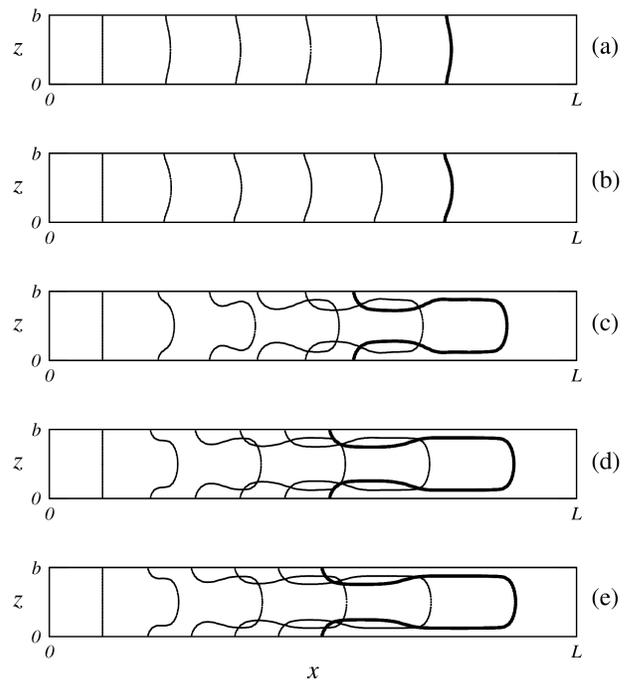}
\caption{Interface evolution in the channel thickness direction for varying diffusion strength. 
Time interval between interfaces is $\delta t \simeq 2.17$ in $b/U$ units. The thick profile in each figure corresponds 
to the latest time. Meniscus regime: (a) $D=0.073$ and (b) $D=0.049$. 
Finger regime : (c) $D=0.024$, (d) $D=0.012$ and (e) $D=0.009$. 
\label{fig:interPe}}
\end{figure}
We first consider two fluids with equal viscosities and densities. 
The size of the interface is set to $\xi=0.57$, which has been previously verified 
to give sufficiently accurate results for the variation of $\phi$ and its spatial derivatives 
across the interface\cite{pagonabarraga}. 
Starting from a flat interface  configuration, we perform a set of five runs at fixed forcing, viscosity and surface tension.  
For each run we choose a different diffusion coefficient, which we fix through the mobility.   
In terms of dimensionless numbers this corresponds to fix $Ca$ and vary $Pe$. 
Parameter values are $U=5\times 10^{-3}$, $\eta=10^{-1}$ and $\sigma=4.6\times10^{-3}$, where $U$ is the expected 
leading interface velocity, calculated as $U=b^2\rho g /(8\eta)$.  Channel dimensions are $b=23$ and $L=500$.
 
In Fig.~\ref{fig:interPe} we show a time sequence of the interface position for each run.  
In our simulations, $v_y=0$ so a flat leading interface 
is located at $z=b/2$. Sequences (a) and (b) correspond to runs with the highest diffusion coefficients.  
In both cases a steady meniscus is clearly observed.  It is also appreciable that the meniscus in sequence~(a),  
corresponding to the highest diffusivity, is less curved than the meniscus in sequence~(b).  
The next three sequences, (c), (d) and (e), show an abrupt change in the interface configuration. 
Instead of a meniscus, we observe a penetrating structure that emerges from the center of the channel 
leaving a thin film of fluid adhered to the solid plates.  The finger width in runs (c)-(e) is 
approximately 17 lattice spacings. For the size of the interface used, the order parameter saturates to 
its equilibrium value at the solid surface.   Nonetheless, to rule out any effects associated to 
the size of the interface,  we have verified that the finger width~(relative to the channel thickness) 
does not depend on $b$, as we will see bellow.
 
All runs achieve a steady state in which the velocity of the leading interface is constant.  
This velocity is the same for runs (a) and (b) and due to mass conservation is slightly larger 
(a few percent) for runs (c), (d) and (e).  The capillary number is not affected much by this 
effect, and we will take it as constant.  The relevant effect is associated to the variation of diffusivity. 

The velocity of the contact line increases with increasing diffusivity, as can be 
deduced from the contact line position in sequences (c), (d) and (e).  Nevertheless, the velocity of the leading 
interface and the width of the penetrating finger are the same for all three runs.   This is a direct confirmation 
of the fact that contact line dynamics are decoupled from leading interface dynamics
in the presence of a thin film, as proposed by Park and Homsy in Ref.\cite{parkhomsy}.

It is clear from these runs that the crossover for penetration is set by the 
difference between the leading interface velocity, $U$, and the
slip velocity at the contact line, $v_s$.  For a meniscus, $v_s = U$, while penetration occurs whenever $v_s < U$.
As $v_s$ depends on the strength of diffusivity,  we can draw as a conclusion that penetration can be achieved by 
increasing $Pe$.   

We now explore the effect of capillarity on the dynamics of the interface.  To do so
we force the interface at fixed velocity, diffusivity and surface tension(resp. viscosity) while 
we vary the viscosity~(resp. surface tension).  As a consequence, 
$Pe$ is fixed while $Ca$ is varied.  

Results are summarized  in Table~\ref{tab:data}.  The first column 
shows parameters for runs in which the viscosity is varied.  We observe that  penetration occurs 
as $\eta$ increases.  The second column in Table~\ref{tab:data} shows results for varying surface tension.   
We observe that penetration occurs as $\sigma$ is decreased.  We can conclude that capillarity plays a 
similar role as diffusivity, as penetration occurs for low values of $Ca$.
\begin{table}
\caption{Parameter values for $\eta$ and $\sigma$ varying runs, 
$U \simeq 5\times 10^{-3}$, $D=7.5\times 10^{-2}$.
\label{tab:data}}
\centering
\begin{ruledtabular}
\begin{tabular}{llll}
\multicolumn{2}{l} {$\sigma=4.6\times10^{-3}$} & \multicolumn{2}{l}{$\eta=10^{-1}$}\cr\hline
$\eta$ & shape &$\sigma$ & shape\cr
0.1&meniscus&0.0044&meniscus\cr
0.2&finger&0.0037&meniscus\cr
0.4&finger&0.0032&meniscus\cr
0.6&finger&0.0027&finger\cr
\end{tabular}
\end{ruledtabular}
\end{table}

\subsection{The Onset of Penetration}
\label{sec:Disc}
Our results suggest that the crossover from the meniscus regime to the thin film regime is controlled at least by 
two mechanisms.  On the one hand, viscous stresses deform the interface.  As a result surface tension tends 
to restore the interface shape to its equilibrium value.  On the other hand, advection causes order parameter gradients. 
As a consequence, diffusivity generates a slip velocity at the contact line.  In this section we will see that the 
balance between these mechanisms is controlled by $Pe$ and $Ca$. 

Let us write the force balance per unit volume of fluid in the frame of reference of the interface.  
We introduce orthogonal curvilinear coordinates, $s$, the arclength along the curve $\phi=0$, and $u$, 
the normal distance to a point on this curve.  In terms of these coordinates the normal component of Eq.~(\ref{eq:ns1}) (in absence of inertial terms) is
\begin{equation}
\partial_u P= - \phi \partial_u \mu + \eta \nabla^2 v_n + \rho g_n,
\label{eq:fb1}
\end{equation}
where the subscript $n$ stands for the normal component and the subscript $u$ denotes differentiation
with respect to $u$.

The force per unit area acting on the interfacial region is obtained 
by integrating~(\ref{eq:fb1}) across the interface: 
\begin{equation}
\Delta P  = -\sigma (\kappa^D_\sigma-\kappa^E_\sigma) + \left(\eta \nabla^2 v_n +  \rho g_n\right)\xi,
\label{eq:fb}
\end{equation}
where the term $\sigma (\kappa^D_\sigma-\kappa^E_\sigma)$ arises from the integration of the chemical potential term\cite{bray},  
with $\kappa^D_\sigma$ and $\kappa^E_\sigma$ being the dynamic and equilibrium curvatures, which are positive for a bump protruding 
in the $x$ direction.   We have assumed that neither of the last two terms in the right hand side vary appreciably across the interface.  
Eq.~(\ref{eq:fb}) should be interpreted as the usual Gibbs-Thomson condition plus a dynamic term proportional to $\xi$, which 
vanishes either in equilibrium or in the sharp interface limit.  

We will now examine Eq.~(\ref{eq:fb}) in the vicinity of the contact line.  
The mesoscopic nature of the interface gives rise to a finite size region where diffusion is important. 
This results in a slip velocity, $v_s$, for the contact line.  We now reproduce the scaling argument 
presented in Ref.\cite{Yeomans01} to obtain the diffusion size, $l_D$, and consequently $v_s$. We will subsequently compare the slip velocity 
to the leading interface velocity in terms of $Ca$ and $Pe$, which are parameters that can be linked to experiments.

The slip velocity and the size of the diffusion region fix the magnitude of viscous dissipation in Eq.~(\ref{eq:fb}), 
\begin{equation}
\Delta P  \sim -\sigma  (\kappa^D_\sigma-\kappa^E_\sigma) + \left(\frac{\eta v_s}{l_D^2} + \rho g_n\right)\xi.
\label{eq:fbs}
\end{equation}
Since in the contact line region the time variation of the order parameter is
$\partial \phi / \partial t \simeq v_s\Delta \phi /\xi$,  the order parameter variation obeys
\begin{equation}
v_s \frac{\Delta \phi}{\xi} \sim \frac{D\Delta \phi}{l_D^2}.
\label{eq:cds} 
\end{equation}
Using Eq.~(\ref{eq:cds}) to eliminate $l_D$ from Eq.~(\ref{eq:fbs}) we get 
$$\Delta P  \sim -\sigma  (\kappa^D_\sigma-\kappa^E_\sigma) + \frac{\eta v_s^2}{D} + \rho g_n\xi.$$
The last term in this expression is order $\xi$, while the rest of terms are of order $\xi^0$. 
The term in the left hand side is the excess pressure drop caused by the  curvature 
difference, which is small in our simulations.  Neglecting both the pressure gradient and the body force 
we extract the following scaling law for the slip velocity: 
$$v_s^2 \sim \frac{\sigma  (\kappa^D_\sigma-\kappa^E_\sigma)D}{\eta}.$$
The interface curvature is a consequence of the underlying velocity profile, which is set 
by the thickness of the channel. Therefore, the curvature difference scales as
$(\kappa^D_\sigma-\kappa^E_\sigma) \sim a/b$, 
with $a$ being a typical amplitude.  Using this expression and measuring $v_s$ in units of the leading 
interface velocity we arrive at the following expression:
$$\left(\frac{v_s}{U}\right)^2 \sim a Ca^{-1}Pe^{-1}.$$
This indicates that both $Pe$ and $Ca$ control how the slip velocity compares to the 
leading interface velocity.   For a meniscus $v_s=U$, so we arrive at the following condition:  
\begin{equation}
Pe=aCa^{-1}.
\label{eq:r3}
\end{equation}

\begin{figure}[t!]
\includegraphics[width=0.45\textwidth]{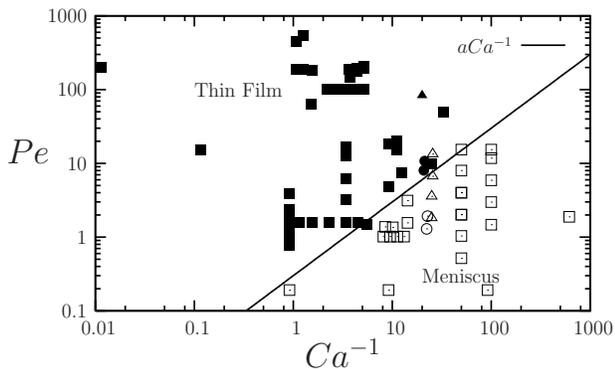}
\caption{
P\'eclet and capillary numbers for meniscus and thin film regimes\label{fig:PhaseDiagram}. 
$\Box$ Meniscus regime. $\blacksquare$ Thin film regime. $\circ$ Meniscus regime (different viscosities). 
$\bullet$ Thin film regime (different viscosities). $\triangle$ Meniscus regime (different densities).  
$\blacktriangle$ Thin film regime (different densities). The solid line corresponds to Eq.~(\ref{eq:r3}), 
with $a \simeq 0.3.$}
\end{figure}
To check the validity of the prediction we perform several runs of forced interfaces varying simulation 
parameters in a wide range(see Tables~\ref{tab:meniscusdata} and~\ref{tab:fingerdata}). 
We cover up to four decades in
the $Pe$ and $Ca$ until numerical stability issues of the code begin to show up.   
In Fig.~\ref{fig:PhaseDiagram} we show a plot of our results in the $Ca^{-1}Pe$ plane.  
Data shown in the figure sketches two regions; at high $Pe$ and $Ca$ values the thin film regime is observed, 
whereas the meniscus corresponds to low values of both parameters.  We also show our prediction, 
for which we find a fitting  value for $a$, namely $a\simeq 0.3.$

Typical experiments with molecular liquids correspond to high, $O(10^2)-O(10^3)$, values 
of the P\'eclet number\cite{libchaber,Petitjeans01}, thus falling in the thin film region sketched 
in the diagram.  However, menisci are expected in systems  with a diffuse interface, such as 
colloid-polymer mixtures\cite{Aarts01}.  In terms of experimental parameters, the diffusion length 
scales as $l_D\sim(\eta \xi^2 D/\sigma(\kappa_\sigma^D - \kappa_\sigma^E))^{1/4}$. 
In colloid-polymer mixtures the ratio $(\xi^2/\sigma)^{1/4}$ is about $10^2$ times larger 
than for molecular liquids.  As a consequence, in such systems menisci should be observable 
at relatively high capillary numbers.   For molecular liquids, this effect can be achieved by decreasing 
the system size,  for instance,  
in microfluidic devices, where the typical size of the channel is a few microns\cite{Tabeling01}.  
For such small sizes, the P\'eclet number is $O(1)$, about $10^3$ times smaller than for traditional 
channels.  Hence, the transition from menisci to thin films would be observable in the regime of 
relatively high capillary number, say $Ca$ $O(10^{-1})-O(10^0)$.   
\subsection{Asymmetric Fluids}
\label{subsec:asymm}
We have shown that the ratio between the leading interface and contact line velocities is controlled 
by the interplay between the P\'eclet and capillary numbers.  We expect that this fact holds for fluids 
of either different densities or viscosities.   A forced interface between asymmetric fluids can be destabilized by
virtue of the Rayleigh-Taylor instability, when the more dense fluid displaces the less dense one.   We explore 
situations for which the instability is absent.  To ensure this, we keep the channel thickness below the first
unstable wavelength, given by $l_c=2\pi(\sigma/\Delta\rho g)^{1/2}$\cite{Chandrasekhar01}.

The forcing is set according to $\vec f = 8\eta(\phi) U_{exp}/b^2A(\phi)\hat x$. 
where the local viscosity is set according to the mixing rule $\eta(\phi)=(\eta_2+\eta_1)(1-c\phi/\phi_{eq})/2$, 
characterized by the viscosity contrast $c=(\eta_2 - \eta_1)/(\eta_2 + \eta_1)$, and $U_{exp}$ is the maximum expected 
velocity for a Poiseuille profile.  The $\phi$ dependent part is set as $A(\phi)=1$ if $c\neq 0$ and 
$A=(\phi+\phi_{eq})/\Delta \phi$ otherwise.  In experiments the typical situation is that an effectively inviscid fluid 
displaces a viscous one, which corresponds to $c\rightarrow 1.$  We approach this situation by setting $c=0.9.$  Following the general
convention in the literature, here we define the capillary number as $Ca=\eta_2U/\sigma$.  
For all cases, we consider that both fluids have the same diffusion coefficient.

We have performed a set of runs in which we vary $Pe$ at fixed $Ca$ for fluids with finite      
density or viscosity contrast.  The details of the runs are summarized in Table~\ref{tab:inhom}. 
For each case, both menisci and fingers can be obtained depending on the value of the P\'eclet number.  As expected, penetration occurs for sufficiently high 
$Pe$.  We can conclude that the appearance of a thin film is independent of the Rayleigh-Taylor instability.  
In Fig.~\ref{fig:PhaseDiagram} we plot results of this section in the $PeCa^{-1}$ plane. For fluids of different densities this value is consistent with the symmetric estimate of $a\simeq0.3$.  For fluids of different viscosities the crossover occurs at $a\simeq 0.5$.  Anyhow, the qualitative behavior remains independent of 
the degree of asymmetry between the fluids. 
\begin{table}
\caption{Parameter values for runs of asymmetric fluids.\label{tab:inhom}}
\centering
\begin{ruledtabular}
\begin{tabular}{llll}
\multicolumn{2}{l} {Varying densities} & \multicolumn{2}{l}{Varying viscosities}\cr
\multicolumn{2}{l} {$\sigma=9.2\times10^{-3}$, $U\simeq 4\times 10^{-3}$} & \multicolumn{2}{l}{$\sigma=4.6\times10^{-3}$, $U\simeq 2\times 10^{-3}$}\cr\hline
$D$ & shape & $D$ & shape\cr
0.146&meniscus&0.0488&meniscus\cr
0.098&meniscus&0.0244&meniscus\cr
0.049&finger&0.0122&meniscus\cr
0.024&finger&0.006&meniscus\cr
0.018&finger&0.001&finger\cr
\end{tabular}
\end{ruledtabular}
\end{table}  

\subsection{Steady state finger in the channel thickness direction}
\label{subsec:finger}
\begin{figure}[b]
\includegraphics[width=0.45\textwidth]{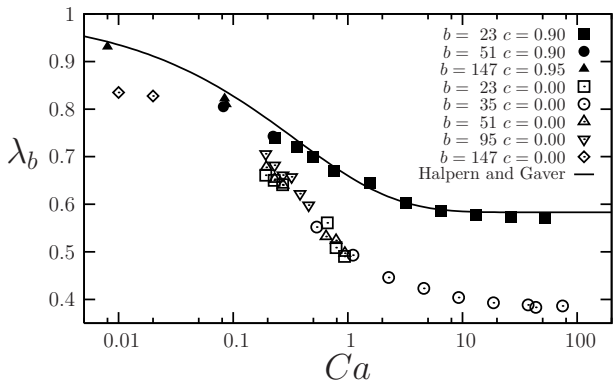}
\caption{Finger width as a function of the $Ca$. The error in the measured finger width 
is calculated from one lattice spacing and corresponds to approximately $\delta \lambda_b \simeq 0.04$ 
in the figure.
\label{fig:lambda_b}}
\end{figure}
We now turn our attention to the steady state finger that appears for high values of the product $CaPe$,  
which is the usual situation in most experiments.  As we have shown in Sec.~\ref{sec:Diff}, diffusion 
only affects the motion of the contact line, and has a negligible effect in the steady state finger.  The 
relevant control parameters, as proposed in the literature, are the capillary number and the viscosity contrast 
between the fluids. The steady state if often characterized by measuring the finger width, $\lambda_b$, which
is the fraction of occupation of fluid~1 relative to the channel thickness. 

We explore $\lambda_b$ as a function of $Ca$ at a given $c$ value.  We consider three situations, $c=0$ and
zero density contrast,  $c=0$ with finite density contrast and $c\neq 0$ with zero density contrast.   For the last 
case we choose $c=0.90$ and $c=0.95$. The low $Ca$ runs have been performed varying $b$ to rule out
lattice artifacts.  We have found that results do not depend on the channel thickness chosen, the 
smallest thickness considered here being $b=23$.   
Tabs.~\ref{tab:statfingerdata},~\ref{tab:statfingerdatanu}  and~\ref{tab:statfingerdatac} display the parameter values used in each run. 

In Fig.~\ref{fig:lambda_b} we plot $\lambda_b$ as a function of $Ca$.  
We find that $\lambda_b$ depends on the viscosity contrast, 
as the $c=0$ points fall in a clearly different curve than the $c=0.9$ 
and $c=0.95$ points.  

On the contrary, the density contrast does not play a relevant role. 
Points belonging to the $c=0$ curve were obtained using five different gap sizes; 
$b=147$, $b=23$, $b=35$, $b=51$ and $b=95$, of which the $b=35$ set was done 
with fluids of different densities.  Results show no difference between 
zero or finite density difference.  In fact, the gap size for the latter 
was large enough for the Rayleigh-Taylor instability to be present.  
This means that the finger can develop as a consequence of low diffusion 
or by virtue of the Rayleigh-Taylor instability.  Still, the steady state 
remains insensitive to the mechanism that leads to penetration and is selected 
by $Ca$ and $c$. 

Previous analytic predictions correspond to the low $Ca$ regime at $c=1$ and were 
carried out first by Bretherton\cite{Bretherton01}, who found that the finger width decays as
$\lambda_b\rightarrow1-1.337Ca^{2/3}$,  as $Ca\rightarrow0$.  
An extension was done by Taylor\cite{Taylor01}, up to  $Ca < 0.09$, for which he reported a 
decaying exponent of one-half.  Numerically, Reinelt and Saffman\cite{reinelt} solved the Stokes 
equations using a finite difference algorithm, and  considered values up to $Ca < 2$, which match the 
one-half exponent of Taylor at small $Ca$.   Halpern and Gaver\cite{Halpern01} extended the prediction beyond $Ca=2$ by means of a  
boundary element analysis  of the Stokes equations. Their results can be fitted to an exponential law
$\lambda_b=1-0.417\left(1-\exp(-1.69Ca^{0.5025})\right)$(shown in Fig.~\ref{fig:lambda_b}), which
reproduces Reinelt and Saffman results and matches the power law prediction of Taylor for low $Ca$.  For large
$Ca$, this law saturates to a limiting value of $\lambda_b=0.583$.   Previous Lattice-Boltzmann
studies have also addressed this problem.   Kang \emph{et al}\cite{Kang01}
studied the range $0.2\leq Ca < 2$, obtaining good agreement with Halpern and Gaver results. 
Langaas and Yeomans\cite{Yeomans03} considered the range $0.079 \leq Ca \leq
4.6$ and were able to reproduce the results of Halpern and Gaver for $Ca < 2$.  For $Ca > 2 $ they
obtained smaller finger widths than those of Halpern and Gaver.  

Our results cover up to five decades 
in the capillary number, $10^{-2} \leq Ca \leq 10^2$\cite{foot}.  They match Halpern and  Gaver
prediction as $c \rightarrow 1$.  Already at $c=0.90$, we reproduce accurately the finger width 
saturation value, for which we find $\lambda_b=0.573 \pm 0.022$.     
For small $Ca$, the error increases for the $c=0.9$ runs.  We improve this situation by increasing the viscosity
contrast, as can be appreciated in Fig.~\ref{fig:lambda_b}.  At $Ca\simeq 0.09$ the error for $c=0.9$ is
4\%,while for $c=0.95$ it reduces to 2\%.  For $Ca=0.008$ the error is $2\%$ at $c=0.95$.   This agreement
shows that the Lattice-Boltzmann approach gives accurate results for a wide range of $Ca$,  improving previous 
results\cite{Yeomans03}.

As can be seen from Fig.~\ref{fig:lambda_b}, fingers with zero viscosity contrast, a case that has not 
been studied previously, are much narrow than fingers with $c=0.9$ or $c=0.95$.  The dependence of the finger 
width with $Ca$ has a power law behavior for $0.1 \leq Ca \leq 1$, with an exponent $m = 0.29\pm0.02$.  
For $Ca$ $O(10)$, the finger width saturates to a notably small value, $\lambda_b\simeq 0.386 \pm 0.014$, 
which remains an open question.  

\begin{figure}[b!]
\includegraphics[width=0.45\textwidth]{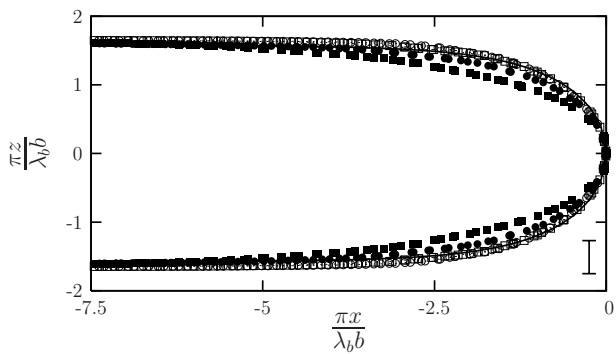}
\caption{Rescaled interface profiles for the lowest and highest $Ca$ values of the $c=0.0$ and $c=0.9$ runs.  
For $c=0.0$: ${\circ} Ca=0.01$ and ${\bullet} Ca=74.68$.  For $c=0.9$: ${\square} Ca=0.17$ and ${\blacksquare}:Ca=52.82$. 
Solid Line: Pitts semi-empirical finger shape.  
The error bar is shown in the bottom-right and corresponds to one lattice spacing.  
The length of the diffuse interface corresponds to approximately one half of a unit in the figure. 
\label{fig:gap_profiles_collapsed.c0.0}}
\end{figure}

We now focus on the shape of the steady finger. 
Finger profiles shown in Fig.~\ref{fig:interPe} are strongly reminiscent of the 
single finger solution of the Saffman-Taylor problem\cite{saffman}, in which fingering 
occurs in the $xy$ plane of a Hele-Shaw cell as a consequence of viscosity or density 
asymmetries between the fluids.  Our problem is fundamentally different because
penetration in the channel thickness can occur for linearly stable 
interfaces in the context of hydrodynamic stability, \textit{e.g.,}  by virtue of low diffusivity 
at the contact line.  Moreover, even in the case where the interface is 
linearly unstable, it is due to a Rayleigh-Taylor instability and not through the
Saffman-Taylor one.  

Still, we compare our finger profiles with the Saffman-Taylor ones. To do so, we recall the 
semi-empiric shape found by Pitts\cite{Pitts01}, which for our geometry reads
\begin{equation}
\cos\left(\frac{\pi z^\prime}{\lambda_b b}\right)=\exp\left(\frac{\pi x^\prime}{\lambda_b b}\right),
\label{eq:Pitts}
\end{equation}
where $x^\prime$ and $z^\prime$ measure the distance from the finger tip. For both $c=0$ and $c=0.9$, Eq.~(\ref{eq:Pitts}) is a good 
approximation to our profiles at low $Ca$. 
This agreement is lost gradually as $Ca$ increases.  In Fig.~\ref{fig:gap_profiles_collapsed.c0.0} 
we show interface profiles for $c=0.0$ and $c=0.9$ at the smallest and largest $Ca$ considered. 
Profiles for all other $Ca$ values lie between the shown profiles and are omitted from the figure. 
For $c=0.0$, Eq.~(\ref{eq:Pitts}) describes better our profiles for large $Ca$, while for $c=0.9$ 
deviations from Pitts result are observed as $Ca$ increases. 

\section{Conclusions}
\label{sec:Conclusions}
We have studied the forced motion of a fluid-fluid interface in  a 
three-dimensional channel by means of a mesoscopic model that takes into 
account contact line dynamics. 

Our results describe two possible scenarios regarding interface dynamics.  
A meniscus regime is found whenever the contact line velocity is comparable 
to the leading interface velocity. Conversely, when the contact line velocity 
is smaller than the leading interface velocity the meniscus configuration is lost, leading 
to penetration of one fluid into the other in a fingering fashion.   A thin film of displaced 
fluid is hence left adhered to the plates of the channel.  

The crossover from the meniscus to the thin film regime is controlled by the competition 
between surface and viscous stresses, as well as by the competition 
of diffusive and advective timescales, on top of the usual hydrodynamic 
instabilities.  These mechanisms can be accounted for through simple scaling arguments. We find a prediction for the crossover in terms of the capillary and P\'eclet numbers which describes 
accurately our numerical results.  Menisci are found for low capillary and P\'eclet numbers, when 
surface tension and diffusion dominate over viscous stresses and advection respectively.  An example of a system where 
diffusion is important is that of colloid-polymer mixtures.  For such systems, 
the relatively large size of the interface together with low surface tensions leads to large diffusion regions near de contact line.  
For instance, in Ref.\cite{Aarts01}
the size of the interface is typically $\xi\simeq 10~\mu $m while the surface tension is $\sigma \simeq 1 \mu$N.  For molecular liquids the 
size of the interface is of the order of nanometers, while $\sigma \simeq 10~$mN$/m$.   As explained in Sec.\ref{sec:Disc}, 
the size of the diffusion region scales as $l_D \sim (\xi^2/\sigma)^{1/4}$, all other parameters kept constant.  
With these values the ratio of the diffusion length between colloid-polymer mixtures and molecular liquids 
is at least two orders of magnitude.  Thin films are obtained for high values of both parameters, when advection and 
viscous stresses are dominant. Our prediction works for both symmetric and asymmetric fluids.  From the 
crossover prediction, we propose that thin films can be assured as long as $CaPe > 0.3$ for symmetric fluids 
and $CaPe > 0.5$ for asymmetric fluids. Anyhow, this values are consistent with the typical 
experimental regime. For instance, experiments of Ref.\cite{libchaber}
were done in a channel of thickness $b=7.95\times10^{-4}$~m, with a silicone oil with  
$\eta=9.3$~cP, $\sigma=20.1$~dyn/cm and $D = 1.4946\times10^{-6}$~cm$^2/$s\cite{Reid01}.  
Typical velocities in the experiments ranged from $U=0.01$~m/s to $U=0.1$~m/s.  
With these values, we obtain $CaPe \simeq 2.6\times10^{4}$, which is consistent with our 
prediction for the thin film regime. 
 
We have examined the steady state of the thin film regime.  We have found, in agreement with  
Ref.\cite{parkhomsy}, that contact line dynamics does not affect the steady state finger shape.  The 
capillary number and the viscosity contrast between the fluids determine the shape of the finger.  

For a low-viscosity fluid pushing a high-viscosity one, the finger narrows with increasing capillary number 
down to a limiting value of $0.57$ in units of the channel thickness. These results agree with the numerical 
results of Ref.\cite{Halpern01} for the wide range of capillary numbers considered.  Due to 
computational limitations we do not investigate fingers with very low capillary numbers.  Nonetheless, as 
we recover results from Ref.\cite{Halpern01}, we expect that the low capillary number limit can be recovered 
by our method as well.

For fluids with equal viscosities we have found a curve of the finger width as a function 
of the capillary number that does not follow any previous results. The width of the finger 
decreases with increasing capillary number, an expected observation, but to a remarkably limiting width of $0.38$ in units of the channel thickness. This contrasts with the saturation value of the asymmetric case.  

The steady state is independent of whether or not the interface is linearly unstable to the Rayleigh-Taylor instability.  This reinforces the 
conjecture of the steady state being independent of the mechanism that first leads to penetration of one fluid to the other. For low 
capillary numbers the shape of our fingers is universal and is consistent with the finger shape of Pitts, which suggests 
that the steady state can be described on simple mechanical equilibrium grounds.  

In a future work we will address the problem of viscous fingering allowing for a non-flat leading interface.  
As we have shown, the shape of the interface across the channel thickness can be controlled by tuning the diffusion strength in the contact line.  Hence, it is possible to describe situations in which the leading interface undergoes a fingering process 
both in the presence and absence of a thin film in the channel thickness direction.  In the presence of a thin film, additional control over the interface shape can be gained by choosing between fluids of equal or different viscosities.   
These features are very convenient to study in detail the three-dimensional effects that arise in the 
viscous fingering problem. 

\section{Acknowledgments}
We acknowledge financial support from Direcci\'on General de Investigaci\'on (Spain) under projects FIS\ 2006-12253-C06-05 and FIS\ 2005-01299.  
R.L.-A. wishes to acknowledge support from CONACyT (M\'exico) and Fundaci\'on Carolina(Spain).  
Part of the computational work herein was carried on in the MareNostrum Supercomputer at Barcelona Supercomputing Center.  
\label{sec:Acknowl}

\appendix
\section{Parameter Values}

\label{sec:Param}
\begin{table}[ht!]
\caption{Parameter values which develop a meniscus. For all runs $b=23$. \label{tab:meniscusdata}}
\begin{ruledtabular}
\centering
\begin{tabular}{cccccc}
$\sigma$   &   $D$    &   $\eta$    &   $U$   &      $Ca$   &   $Pe$\cr\hline
0.0092   &   0.0976   &   0.100   &   0.0067   &    0.0723   &   1.56\cr
0.0092   &   0.0488   &   0.100   &   0.0067   &    0.0724   &   3.12\cr
0.0092   &   0.0976   &   0.100   &   0.0022   &    0.0239   &   0.52\cr
0.0092   &   0.0488   &   0.100   &   0.0022   &    0.0239   &   1.04\cr
0.0092   &   0.0244   &   0.100   &   0.0021   &    0.0232   &   2.02\cr
0.0092   &   0.0122   &   0.100   &   0.0021   &   0.0229   &   3.98\cr
0.0092   &   0.0244   &   0.100   &   0.0021   &    0.0232   &   2.00\cr
0.0092   &   0.0122   &   0.100   &   0.0021   &    0.0229   &   3.98\cr
0.0092   &   0.0066   &   0.100   &   0.0021   &    0.0227   &   7.86\cr
0.0092   &   0.0031   &   0.100   &   0.0021   &   0.0223   &   15.5\cr
0.0092   &   0.0976   &   0.015   &   0.0063   &    0.0102   &   1.48\cr
0.0092   &   0.0488   &   0.015   &   0.0063   &    0.0103   &   2.97\cr
0.0092   &   0.0244   &   0.015   &   0.0063   &    0.0103   &   5.94\cr
0.0092   &   0.0122   &   0.015   &   0.0063   &    0.0103   &   11.9\cr
0.0092   &   0.0091   &   0.015   &   0.0063   &    0.0103   &   15.8\cr
0.0092   &   0.0091   &   0.015   &   0.0010   &    0.0016   &   2.52\cr
0.0009   &   0.0240   &   5.000   &   0.0002   &    1.0963   &   0.20\cr
0.0009   &   0.0240   &   0.500   &   0.0002   &    0.1096   &   0.20\cr
0.0009   &   0.0240   &   0.050   &   0.0002   &    0.0110   &   0.20\cr
0.0044   &   0.0750   &   0.100   &   0.0033   &    0.0755   &   1.02\cr
0.0037   &   0.0750   &   0.100   &   0.0033   &    0.0889   &   1.02\cr
0.0032   &   0.0750   &   0.100   &   0.0033   &    0.1046   &   1.02\cr
0.0027   &   0.0750   &   0.100   &   0.0033   &    0.1230   &   1.02\cr
0.0044   &   0.0750   &   0.100   &   0.0043   &    0.0984   &   1.32\cr
0.0037   &   0.0750   &   0.100   &   0.0043   &    0.1158   &   1.32\cr
\end{tabular}
\end{ruledtabular}
\end{table}

\begin{table}[ht!]
\caption{Parameter values which develop a finger.  For all runs $b=23$. \label{tab:fingerdata}}
\begin{ruledtabular}
\centering
\begin{tabular}{cccccc}
$\sigma$   &   $D$    &   $\eta$    &   $U$   &      $Ca$   &   $Pe$\cr\hline
0.0092   &   0.0244   &   0.1   &   0.0080   &     0.09   &   7.54\cr
0.0092   &   0.0122   &   0.1   &   0.0080   &     0.09   &   15.08\cr
0.0092   &   0.0092   &   0.1   &   0.0080   &     0.09   &   20.10\cr
0.0092   &   0.0092   &   0.1   &   0.0040   &     0.04   &   10.06\cr
0.0092   &   0.0015   &   0.1   &   0.0030   &     0.03   &   45.24\cr
0.0092   &   0.0976   &   0.4   &   0.0067   &     0.29   &   1.56\cr
0.0092   &   0.0488   &   0.4   &   0.0067   &     0.29   &   3.16\cr
0.0092   &   0.0244   &   0.4   &   0.0067   &     0.29   &   6.32\cr
0.0092   &   0.0122   &   0.4   &   0.0067   &     0.29   &   12.64\cr
0.0092   &   0.0092   &   0.4   &   0.0067   &      0.29   &   16.84\cr
0.0046   &   0.0122   &   0.1   &   0.0050   &     0.11   &   4.72\cr
0.0046   &   0.0061   &   0.1   &   0.0050   &     0.11   &   18.86\cr
0.0046   &   0.0732   &   0.2   &   0.0050   &     0.22   &   1.58\cr
0.0046   &   0.0732   &   0.4   &   0.0050   &     0.43   &   1.58\cr
0.0046   &   0.0732   &   0.6   &   0.0050   &     0.65   &   1.58\cr
0.0046   &   0.0732   &   0.8   &   0.0050   &     0.87   &   1.58\cr
0.0005   &   0.0073   &   0.8   &   0.0050   &     8.70   &   15.72\cr
0.0000   &   0.0008   &   0.8   &   0.0050   &     86.98   &  157.10\cr
0.0011   &   0.0183   &   0.8   &   0.0125   &     8.70   &   15.72\cr
0.0000   &   0.0008   &   0.8   &   0.0050   &     86.98   &  157.10\cr
0.0009   &   0.0240   &   1.0   &   0.0010   &     1.10   &   0.96\cr
0.0009   &   0.0192   &   1.0   &   0.0010   &     1.10   &   1.20\cr
0.0009   &   0.0144   &   1.0   &   0.0010   &     1.10   &   1.60\cr
0.0009   &   0.0120   &   1.0   &   0.0010   &     1.10   &   1.92\cr
0.0091   &   0.0960   &   1.0   &   0.0100   &     1.10   &   2.40\cr
0.0009   &   0.0060   &   5.0   &   0.0002   &     1.10   &   0.76\cr
0.0009   &   0.0012   &   5.0   &   0.0002   &    1.10   &   3.84\cr
0.0027   &   0.0750   &   0.1   &   0.0048   &     0.18   &   1.48\cr
\end{tabular}
\end{ruledtabular}
\end{table}

\begin{table}[ht!]
\caption{Parameter values for stationary fingers presented in Sec.~\ref{subsec:finger}.
\label{tab:statfingerdata}}
\begin{ruledtabular}
\centering
 \begin{tabular}{ccccccc}
 $\sigma$ & $\eta$ & $U$ & $Ca$ & $D$ & $Pe$ & $\lambda_b$ \cr\hline 
 \multicolumn{7}{l}{Uniform Forcing, $c=0.0$}\cr
 \multicolumn{7}{l}{$b=23$}\cr\hline
 0.0044 & 0.10 & 0.0084 & 0.192 & 0.00185 & 104.2 & 0.661\cr
 0.0037 & 0.10 & 0.0085 & 0.228 & 0.00222 & 87.87 & 0.650\cr
 0.0032 & 0.10 & 0.0085 & 0.270 & 0.00265 & 73.95 & 0.641\cr
 0.0027 & 0.20 & 0.0089 & 0.662 & 0.00064 & 321.1 & 0.561\cr
 0.0023 & 0.20 & 0.0090 & 0.790 & 0.00076 & 272.34 & 0.509\cr
 0.0019 & 0.20 & 0.0091 & 0.936 & 0.00091 & 229.33 & 0.490\cr
 \multicolumn{7}{l}{$b=51$}\cr\hline
 0.0044 & 0.10 & 0.0085 & 0.194 & 0.00442 & 97.68 & 0.678\cr
 0.0037 & 0.10 & 0.0085 & 0.230 & 0.00449 & 96.84 & 0.653\cr
 0.0032 & 0.10 & 0.0086 & 0.272 & 0.00459 & 95.59 & 0.640\cr
 0.0027 & 0.20 & 0.0087 & 0.646 & 0.00490 & 90.31 & 0.532\cr
 0.0023 & 0.20 & 0.0090 & 0.789 & 0.00489 & 93.75 & 0.524\cr
 0.0019 & 0.20 & 0.0091 & 0.941 & 0.00490 & 95.05 & 0.497\cr
 \multicolumn{7}{l}{$b=95$}\cr\hline
 0.0044 & 0.10 & 0.0084 & 0.192 & 0.01608 & 49.64 & 0.706\cr
 0.0037 & 0.10 & 0.0085 & 0.229 & 0.01621 & 49.75 & 0.682\cr
 0.0032 & 0.10 & 0.0086 & 0.272 & 0.01634 & 49.85 & 0.661\cr
 0.0027 & 0.10 & 0.0087 & 0.323 & 0.01648 & 49.93 & 0.658\cr
 0.0023 & 0.10 & 0.0087 & 0.383 & 0.01661 & 50 & 0.622\cr
 0.0019 & 0.10 & 0.0088 & 0.455 & 0.01675 & 50.04 & 0.598\cr
 \multicolumn{7}{l}{$b=147$}\cr\hline
 0.0046 & 0.005 & 0.0095 & 0.010 & 0.1464 & 9.54 & 0.835\cr
 0.0046 & 0.01  & 0.0094 & 0.020 & 0.1464 & 9.54 & 0.828\cr\hline
\end{tabular}

\end{ruledtabular}
\end{table}

\begin{table}[ht!]
\caption{Parameter values for stationary fingers presented in Sec.~\ref{subsec:finger}. For all runs $b=35$. 
\label{tab:statfingerdatanu}}
\begin{ruledtabular}
\centering
\begin{tabular}{ccccccc}
 $\sigma$ & $\eta$ & $U$ & $Ca$ & $D$ & $Pe$ & $\lambda_b$ \cr\hline 
 \multicolumn{7}{l}{Non-Uniform Forcing, $c=0.0$}\cr
 \multicolumn{7}{l}{$b=35$}\cr\hline
 0.0046 & 0.50 & 0.0049 & 0.535 & 0.00071 & 241.17 & 0.552\cr
 0.0023 & 0.50 & 0.0051 & 1.109 & 0.00071 & 250.09 & 0.493\cr
 0.0011 & 0.50 & 0.0052 & 2.270 & 0.00071 & 255.98 & 0.450\cr
 0.0006 & 0.50 & 0.0053 & 4.604 & 0.00071 & 259.57 & 0.423\cr
 0.0003 & 0.50 & 0.0053 & 9.272 & 0.00071 & 261.38 & 0.404\cr
 0.00014 & 0.50 & 0.0054 & 18.617 & 0.00071 & 262.39 & 0.393\cr
 0.00007 & 0.50 & 0.0054 & 37.306 & 0.00071 & 262.91 & 0.388\cr
 0.00002 & 0.50 & 0.0016 & 43.688 & 0.00037 & 154.85 & 0.383\cr
 0.00004 & 0.50 & 0.0054 & 74.685 & 0.00071 & 263.16 & 0.386\cr\hline
\end{tabular}
\end{ruledtabular}
\end{table}

\begin{table}[ht!]
\caption{Parameter values for stationary fingers presented in Sec.~\ref{subsec:finger}.
\label{tab:statfingerdatac}}
\begin{ruledtabular}
\centering
\begin{tabular}{ccccccc}
$\sigma$ & $\eta_2$ & $U$ & $Ca$ & $D$ & $Pe$ & $\lambda_b$ \cr\hline 
 \multicolumn{7}{l}{Uniform Forcing, $c=0.9$}\cr
 \multicolumn{7}{l}{$b=23$}\cr\hline
 0.0046 & 0.38 & 0.0092 & 0.756 & 0.00976 & 21.57 & 0.670\cr
 0.0011 & 0.38 & 0.0097 & 3.218 & 0.00244 & 91.76 & 0.603\cr
 0.0023 & 0.38 & 0.0095 & 1.569 & 0.00488 & 44.73 & 0.644\cr
 0.0006 & 0.38 & 0.0099 & 6.519 & 0.00122 & 185.91 & 0.585\cr
 0.0003 & 0.38 & 0.0099 & 13.124 & 0.00061 & 374.28 & 0.578\cr
 0.0001 & 0.38 & 0.0100 & 26.373 & 0.00031 & 752.09 & 0.574\cr
 0.0001 & 0.38 & 0.0100 & 52.817 & 0.00015 & 1506.22 & 0.572\cr\hline
 \multicolumn{7}{l}{$b=51$}\cr\hline
 0.0046 & 0.095 & 0.0040 & 0.082 & 0.0061 & 33.3 & 0.805\cr
 0.0046 & 0.095 & 0.0108 & 0.223 & 0.0061 & 90.3 & 0.743\cr
\multicolumn{7}{l}{Uniform Forcing, $c=0.95$}\cr
  \multicolumn{7}{l}{$b=147$}\cr\hline
 0.0046 & 0.095 & 0.0041 & 0.084 & 0.14 & 4.07 & 0.822\cr
 0.0046 & 0.049 & 0.0007 & 0.008 & 0.14 & 0.71 & 0.931\cr
 0.0046 & 0.488 & 0.0008 & 0.087 & 0.01 & 9.89 & 0.810\cr
\end{tabular}
\end{ruledtabular}
\end{table}

\newpage

\clearpage

\clearpage

\clearpage

\clearpage


\begin{thebibliography}{10}
\bibitem{foot}
Due to numerical stability issues, the diffusion coefficient, $D$, cannot be
  set to arbitrarily small values. For that reason, runs for $Ca < 10^{-1}$
  were done at relatively low $CaPe$. To obtain a stable finger in this regime
  we modified the initial condition to be an infinitely long finger of fluid~1
  penetrating fluid~2, thus eliminating contact line effects.
\end{thebibliography}

\end{document}